# Design Space Exploration to Find the Optimum Cache and Register File Size for Embedded Applications

**Mehdi Alipour** [1], **Mostafa E. Salehi**[1], **Hesamodin shojaei baghini**[2]
[1] Islamic Azad University, Qazvin Branch, Qazvin 34185-1416 Iran.
[2] Computer Engineering Department, Javid University of Jiroft, Azadi Ave, Jiroft, Iran.

**Abstract -** *In the future, embedded processors must process more computation-intensive network applications and internet traffic and packet-processing tasks become heavier and sophisticated. Since the processor performance is severely related to the average memory access delay and also the number of processor registers affects the performance, cache and register file are two major parts in designing embedded processor architecture. Although increasing cache and register file size leads to performance improvement in embedded applications and packet-processing tasks in high traffic networks with too much packets, the increased area, power consumption and memory hierarchy delay are the overheads of these techniques. Therefore, implementing these components in the optimum size is of significant interest in the design of embedded processors. This paper explores the effect of cache and register file size on the processor performance to calculate the optimum size of these components for embedded applications. Experimental results show that although having bigger cache and register file is one of the performance improvement approaches in embedded processors, however, by increasing the size of these parameters over a threshold level, performance improvement is saturated and then, decreased.*

## 1　Introduction

In recent years embedded application and internet traffic become heavier and sophisticated so, future embedded processors will be encountered by more computation-intensive embedded applications, in this way, designing high performance processors is recommended. By scaling down the feature size, technology and presentation of chip multiprocessors (CMP) that are usually multi-thread processors, somehow the user's performance necessity have guaranteed. Inseparable parts in designing these processors are cache and register file because the performance of a processor is severely related to cache access and also having enough registers.

Recently in numerous researches, multi-thread processors are used to design a fast processor especially in network processors [4], [9], [11], [23], [25], and [26]. In [3] a Markov model based on fine grain multithreading is implemented. Analytical Markov model is faster than simulation and has dispensable inaccuracy. In this chain, stalled threads defined as states and transitions are based on cache contention between threads.

Cache memories are usually used to improve the performance and power consumption by bridging the gap between the speed and power consumption of the main memory and CPU. Therefore, the system performance and power consumption is severely related to the average memory access time and power consumption which makes cache as a major part in designing embedded processor architectures.

In [4] cache misses are introduced as a factor for reducing memory level parallelism between threads. In [5] thread criticality prediction has been used and for better performance, resources are given to the threads that have higher L2 cache misses which are called the most critical threads.

To improve packet-processing in network processors, [6]-[8] have applied direct cache access (DCA). In [9] processor architecture is based on the simultaneous multithreading (SMT) and cache miss rate is used to show the performance improvement. To find out the effect of cache access delay on performance, a comparison between multi-core and multi-thread processors has been performed in [10]. Likewise, victim cache is an approach to improve the performance of a multi-thread processor [11].

All recent researches are based on the comparison results with single-core single-thread processors. In the other word multi-thread processors are the heir of the single thread processors [23], [25], and [26]. Hence, evaluating the effective parameters like cache and register file sizes are required for designing a multithread processor. The basic purpose of this paper is to study the effect of the cache size on the performance because embedded processors process computation and data intensive applications and larger cache sizes will present better performance.

Generally, one of the easiest way to improve the performance of embedded and network processors is increasing the cache size [2], [12], [13], [14], and [22]-[26] but this improvement, severely increase the occupied area and power consumption of the processor. So, it is necessary to find a cache size that creates tradeoffs between performance and power-area of the processor.

From other point of view, because of the performance per area parameter, higher performance in a specified area budget is one of the most important needs of a high performance embedded processor. A negative point of the recent researches is that they don't have any constraints on the cache size.

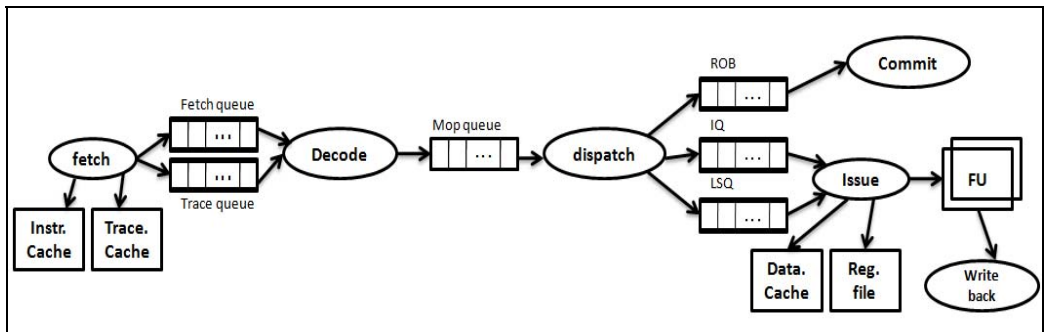

Fig1. Processor pipeline of Multi2sim simulator [19].

Because of the limited area budget in embedded processors, in this paper we have found the optimum size of L1 and L2 cache and also, because of the longer latency of bigger caches, best size of memory hierarchy in relation to this parameter has been calculated.

As mentioned above, another inseparable part in designing embedded processors is register file. Like the cache, size of this parameter has the fundamental effect on the better processor performance. To improve the performance of a embedded processor, a large register file must be implemented. However, larger register files occupy more area and make a worse critical path [18]. Therefore, exploring the optimum size of the register file is the second purpose of this paper. The high importance of this issue is based on the fact that some parameters encourage designer to have a large register file. Generally embedded processors are implemented in multi-issue architecture and out of order (OOO) instruction execution that has renaming logic [16]-[18], [23], [25], and [26]. On the other hand, because register files are shared in multi-thread processors, observing the fact: increment the common parts in design, force the designer to have a larger register file [1]. These parameters also make higher importance for register file size. In [15] effects of register file size in SMT processors have been studied. However, high budget for the number of registers has used. In recent researches effect of register file and cache size in the same time is not studied, so in this paper this issue will be studied too.

## 2 Simulation environment

For simulation, we have used Multi2sim version 2.3.1 [19], a superscalar multi-thread multi-core simulation platform which has 5 stages of pipeline named *fetch, decode, dispatch, issue, writeback, and commit*. This simulator executes x86 instruction sets. Fig.1 shows a block diagram of the processor pipeline modeled in Multi2Sim. In the fetch stage, instructions are read from the instruction or the trace cache. Depending on their origin, they are placed either in the fetch queue or the trace queue. The former contains raw macroinstruction bytes, while the latter stores pre-decoded microinstructions (uops). In the decode stage, instructions are read from these queues, and decoded if necessary. Then, uops are placed in program order into the uop queue. The fetch and decode stages form the front-end of the pipeline [19]. The dispatch stage takes uops from the uop queue, renames their source and destination registers, and places them into the reorder buffer (ROB) and the instruction queue (IQ) or load-store queue (LSQ). The issue stage is in charge of searching both the IQ and LSQ for instructions with ready source operands, which are schedule to the corresponding functional unit or data cache. When and uop completes, the writeback stage stores its destination operand back into the register file. Finally, the completed uops at the head of the ROB are taken by the commit stage and their changes are confirmed. So the commit stage is where we can log and count the number of committed instructions for performance comparison. Detail of this simulation is described in the simulation method and results section.

Multi2sim can run programs in multi issue platform, but to evaluate the requirements of each thread we have used the single issue model for comparison. We changed and compiled the source code of simulator on a 2.4GHz, dual core processor with 4GB of RAM and 6MB of cache that run fedora 10 as an operating system. Base on this configuration the average time of each simulation is about 20 minutes.

## 3 Benchmarks

The aim of this paper is to calculate the optimum cache and register file size. Because embedded applications are so pervasive homogenous applications they cannot be a good choice for DSE. Hence we have applied our DSE on heterogeneous applications, such that in some of them data cache is more important and in the others instruction cache is more important. So we apply PacketBench [20] and MiBench [27] respectively. PacketBench is a good platform to evaluate the workload characteristics of network processors. It reads and writes packets from and to real packet traces, and manages packet memory, and implements a simple application programming interface API. This involves reading and writing trace files and placing packets into the memory data structures used internally by PacketBench. On a network processor, many of these functions are implemented by specialized hardware components and therefore should not be considered part of the application. Programs in this tool are categorized in 3 parts: *1- IP forwarding* which is corresponding to current internet standards. *2- Packet classification* which is commonly used in firewalls and monitoring systems. *3- Encryption,* which is a function that actually modifies the entire payload of the packet.

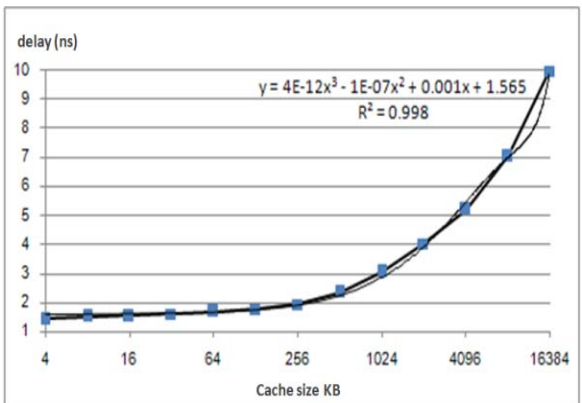

Fig.2. Effect of cache size on cache access delay

Specific applications that respectively we used from each category are IPv4-Lctrie, Flow-Classification and IPSec.IPv4-trie performs RFC1812-based packet forwarding. This implementation is derived from an implementation for the Intel IXP1200 network processor. This application uses a multi-bit trie data structure to store the routing table, which is more efficient in terms of storage space and lookup complexity [20].Flow classification is a common part of various applications such as firewalling, NAT, and network monitoring. The packets passing through the network processor are classified into flows which are defined by a 5-tuple consisting of the IP source and destination addresses, source and destination port numbers, and transport protocol identifier. The 5-tuple is used to compute a hash index into a hash data structure that uses link lists to resolve collisions [20]. IPSec is an implementation of the IP Security Protocol [27], where the packet payload is encrypted using the Rijndael Advanced Encryption Standard (AES) algorithm [28]. This is the only application where the packet payload is read and modified.

MiBench is a combination of six deferent categories. We have selected 3 of them: *1- Dijkstar* from network category, *2- Susan (corners)* from automotive and industrial control category, and *3- String-search* from office category. The Dijkstra benchmark constructs a large graph in an adjacency matrix representation and then calculates the shortest path between every pair of nodes using repeated executions of Dijkstra's algorithm [49]. *Susan* is an image recognition package. It was developed for recognizing corners and edges in magnetic resonance images of the brain [27]. *String-search* searches for given words in phrases using a case insensitive comparison algorithm.

## 4    Simulation method and results

Purpose of this paper is to evaluate optimum size of cache and register file. At first, we describe the methodology to extract proper size of cache. For this purpose, it is necessary to configure the simulator in the way that just the size of cache be the parameter that has affects on the performance. So, for each application the execution number of the main function is calculated in different sizes of L1 and L2 caches. For this purpose we made changes in some parts of simulator source code to calculate the cycles of sending a packet (the cycles that are used to execute the main function of each application).

To calculate the beginning address and the end address of the main function we disassemble the executable code of each benchmark application and extract these addresses and then these parameters are back annotated to commit.c and processor.h file of Multi2sim simulator where a thread is executed.

By these changes we can calculate the number of x86 instructions and macroinstructions and the execution cycles for each specific function.

The second step is to run the simulator with different cache sizes. But the worthwhile point is that although based on the recent researches that recommend doubling the cache size for improving the performance of a processor, however during doubling the cache size, important parameters like area power and cache access delay must be considered. For this purpose we have used CACTI 5.0 [21], a tool from HP that is a platform to extract parameters relevant to cache size considering fabrication technology. Most important parameters that we used in this research are in table 1.

To compare the performance based on the cache size, extracted results from cacti (L1 and L2 cache access delay) are back annotated to Multi2sim. This work has been done by calculating the simulator cycle time and comparing it to the results of cache access time from CACTI. In this way when the cache size is changed, actual cache access delays are considered.

As can be seen in Fig.2, increasing the cache size, leads to more cache access delays.For exploring the cache size, the other simulator parameters are set to the default value, because the purpose is to find the best cache size for a single-thread single-core processor for embedded applications. i.e. width of the pipeline stages must be one (issue-width =1).

Table 1
The most important parameters used in cacti

|  | L1 cache | L2 cache |
|---|---|---|
| Cache size | Variable | Variable |
| Cache line size | Variable | Variable |
| Associatively | Variable | Variable |
| Number of banks | 1 | 1 |
| Technology node (nm) | 90nm | 90nm |
| Read/write ports | 1 | 1 |
| Exclusive read ports | 0 | 0 |
| Exclusive write ports | 0 | 0 |
| Change tag | No | No |
| Type of cache | Fast | normal/serial |
| Temperature (K) | 300-400 | 300-400 |
| RAM cell/transistor type in data array | ITRS-HP | Global |
| RAM cell/transistor type in tag array | ITRS-HP | Global |

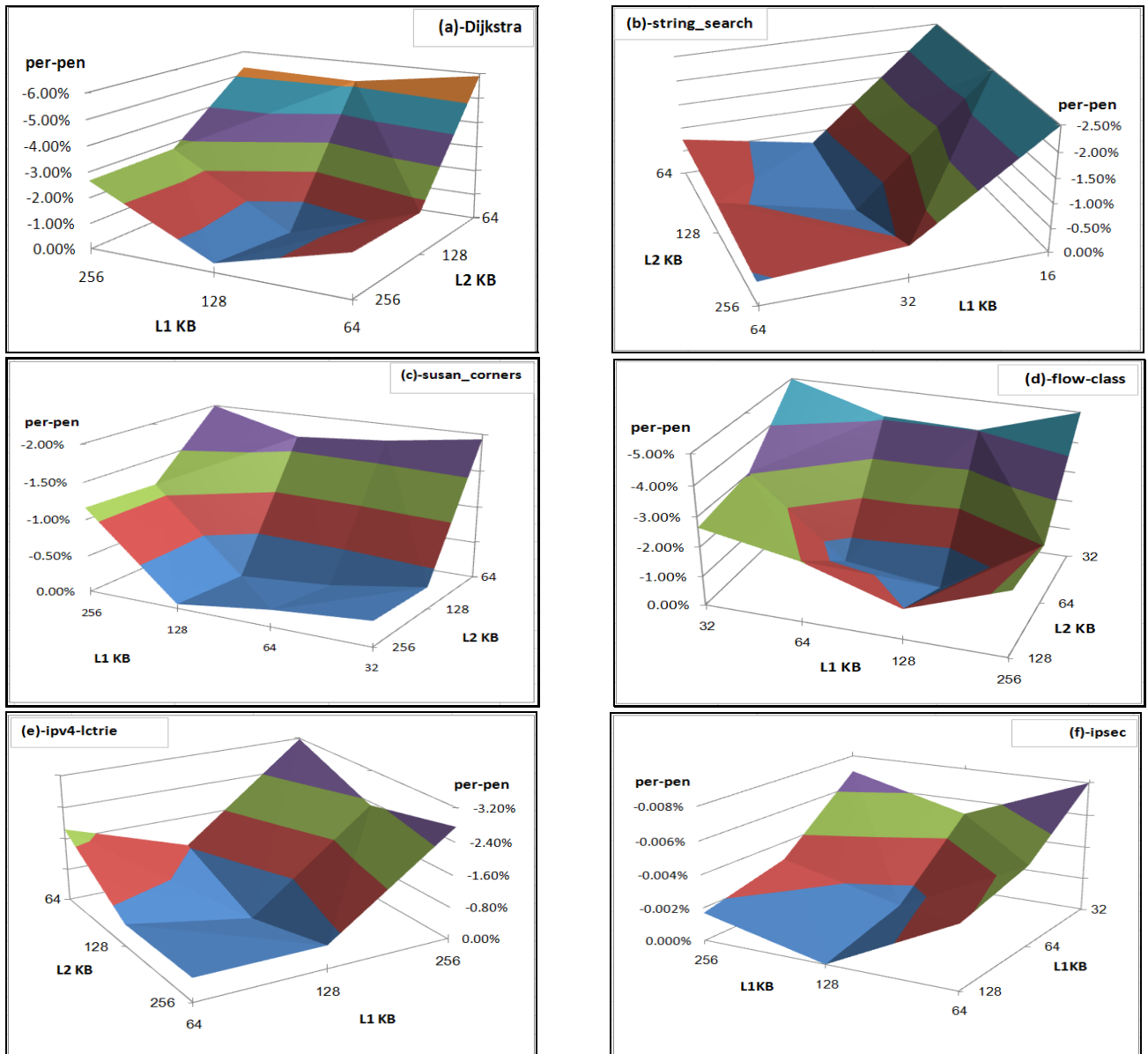

Fig.3. Effect of cache size on the performance (a):Dijkstra, (b): String_search, (c):Susan.corners, (d):flow_class, (e):ipv4_lctrie, (f): ipsec

## 5 Analysis of the simulation results

Fig.3 shows the extracted result of our simulations. In this figure each axis is labelled and the vertical axis (per-pen) shows the performance penalty of related cache size configuration. Based on these results, by increasing the cache size we can achieve more hit rates, however, because of the longer cache access time of larger caches, from a specific point (best cache size) performance improvement is saturated and then even decreased. In other word, doubling the cache size cannot always improve the performance. From other point of view, area budget is limited and always we can't have a large cache, so, considering the sizes smaller and near the best cache size, performance degradations are negligible (3% in average).

To calculate the optimum size of register file, we have applied the parameters used for calculating best cache size, however, to find out just the effect of register file size on the performance, we used the best cache size (L1 and L2) concluded in the previous section for the cache size and run the simulator accordingly. Fig.4 shows the results of this part. In this picture vertical column shows the performance effect (performance penalty) of register file size.

Numbers in this column are relative to the best size of register file. It shows that although for all applications, in average, the best size of register file is 68 and above but in sizes near the half of this size performance penalty is lower that 5%.

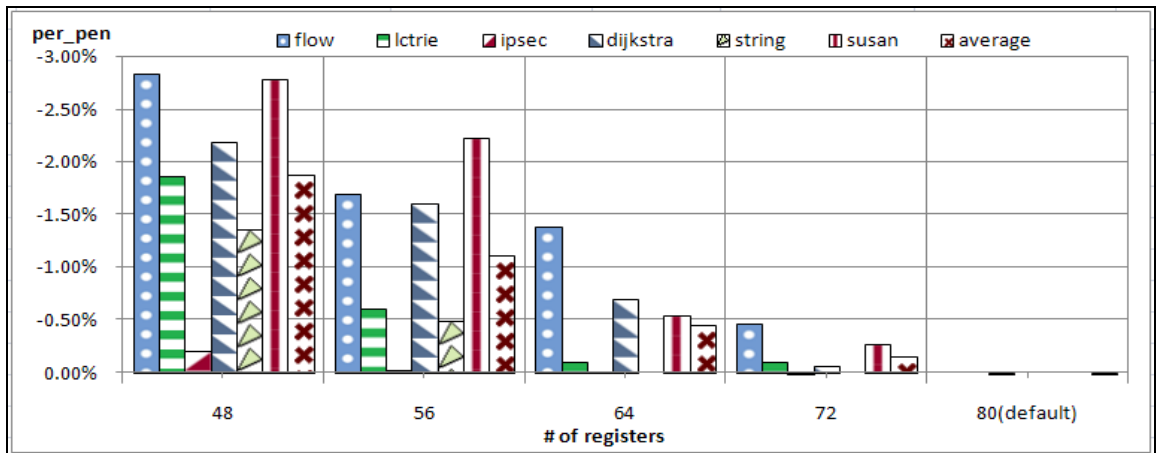

Fig.4. Effect of register file size on performance of different embedded applications.

Also fig.4 shows that reducing the register file size always decrease the performance but sometimes, by doubling the register file size we don't have noticeable performance improvement. So the first point that the highest performance is reached, is introduced as the best size for register file.

It is worthwhile to say that in Fig.4 the effect of cache size and register file size can be seen. In other side of view based on recent researches [23, 25, 26] to have a multi-thread architecture we need more area budget, and to run this architecture in the best performance that can be met, multi issue architecture with renaming logic , ROB, LSQ, IQ and other OOO components which occupy large area budget are needful. Base on these simulations, we calculated 2 point for cache and register file sizes: *1- best size* that has no performance penalty and occupy bigger area budget and *2- optimum size* that has about 3% performance penalty and occupy smaller area budget so, we can deduce that in the optimum size of cache and register file we have saved the area budget of the processors and qualification to run multi-threads in the higher issue widths is obtained. In other word in lower area more performance is achieved and causes to growth the most important parameters for embedded applications that is performance per area.

As mentioned before the multi-thread processors are the heir of single thread processors. So, extracting the best size for important parameters like cache size and register file size is necessary.

## 6 Conclusion

In this paper we have studied the effect of cache and register file size on the performance of an embedded processor and extracted the best size of these two parameters for embedded applications. Simulation results show that for selected benchmarks the best size of L1 and L2 caches are 64KB and 128KB respectively, and the best size of register file is 80. Experiments show that although by increasing the cache size performance will improve, but in a specific point the performance improvement is saturated and then decreased. Also increasing register file size cannot always improve the performance and in a specific size the performance improvement will be saturated. From the area point of view, based on the results of this research, when we select half of the best size of the cache and register file, performance penalty is about 3% in average. In other word in sizes lower than best size the acceptable performance can be met. It means we can reach performance requirements in lower area and also have a better performance/area parameter.